\documentclass[preprint,preprintnumbers,amsmath,amssymb]{revtex4}

\newcommand{\supercite}[1]{\textsuperscript{\cite{#1}}}

\usepackage{amssymb}
\usepackage{graphicx}% Include figure files
\usepackage{bm}% bold math

\begin{document}
%\bibliographystyle{naturemag}

%\preprint{APS/123-QED}

\title{Natural Light Cloaking for Aquatic and Terrestrial Creatures}

\author{Hongsheng Chen$^{1,2,3*}$}
\author{Bin Zheng$^{1,2,3}$}
\author{Lian Shen$^{1,2,3}$}
\author{Huaping Wang$^{4}$}
\author{Xianmin Zhang$^{1,2}$}
\author{Nikolay Zheludev$^{5,6}$}
\author{Baile Zhang$^{6,7*}$}

\affiliation{
$^1$The Electromagnetics Academy at Zhejiang University, Zhejiang University, Hangzhou 310027, China.\\
$^2$Department of Information Science and Electronic Engineering, Zhejiang University, Hangzhou 310027, China.\\
$^3$State Key Laboratory of Modern Optical Instrumentation, Zhejiang University, Hangzhou 310027, China. \\
$^4$Marvell Technology Group Boston, Marlborough, Massachusetts 01752, USA.\\
$^5$Optoelectronics Research Centre and Centre for Photonic
Metamaterials, University of Southampton, Southampton SO17
1BJ,United Kingdom.\\
$^6$Centre for Disruptive Photonic Technologies, Nanyang Technological University, Singapore 637371, Singapore.\\
$^7$Division of Physics and Applied Physics, School of Physical and
Mathematical Sciences, Nanyang Technological University, Singapore
637371, Singapore.\\
\normalsize{$^\ast$To whom correspondence should be addressed;}\\
\normalsize{E-mail:  hansomchen@zju.edu.cn (H.C.);blzhang@ntu.edu.sg(B.Z.)}}


%\begin{abstract}
%\end{abstract}

%\pacs{}% PACS, the Physics and Astronomy
\maketitle

\textbf{A
cloak that can hide living creatures from sight is a common feature of
mythology but still remains unrealized as a practical device. To
preserve the phase of wave, the previous cloaking solution proposed
by Pendry \emph{et al.} required transforming
electromagnetic space around the hidden object in such a way that
the rays bending around it have to travel much faster than those
passing it by. The difficult phase preservation requirement is the
main obstacle for building a broadband polarization insensitive
cloak for large objects. Here, we suggest a simplifying version of
Pendry's cloak by abolishing the requirement for phase preservation
as irrelevant for observation in incoherent natural light with human
eyes that are phase and polarization insensitive. This allows the
cloak design to be made in large scale using commonly available
materials and we successfully report cloaking living creatures, a
cat and a fish, in front of human eyes.}

\newpage

Invisibility cloaking\supercite{alu,leonhardt,pendry,schurig,wenshan_nature,li_carpet,ruopeng,valentine,gabrielli,huifeng,ergin,baile_calcite,xianzhong,edwards,
su_nonsuperluminal,hongsheng_scirep,smith_smartcloak,landy_directional} was almost inconceivable until the ingenious
theory of macroscopic invisibility cloaking was proposed based on
transformation optics principles\supercite{leonhardt,pendry}.
Because of the tremendous difficulty in practical realization,
various
approximations\supercite{schurig,wenshan_nature,li_carpet,xisheng}
were generally taken to simplify the complexity of a perfect cloak.
For example, nonmagnetic optical cloaking was proposed by embedding
metal nanowires in a dielectric material\supercite{wenshan_nature}.
With these approximations, cloaks that could hide objects roughly
one wavelength large (an optical wavelength amounts to the scale of
single-celled organism) have been experimentally demonstrated in
both microwave\supercite{schurig,ruopeng,huifeng} and optical
spectrum\supercite{valentine,gabrielli,ergin}. Shortly, successful
attempts have been made to push the cloaking technology into much
larger scales. The natural anisotropic crystal of calcite has been
used to realize invisibility at the scale of
millimeters\supercite{baile_calcite,xianzhong} for specific
polarized visible light in an environment of optical immersion oil.
Developing a method of natural light cloaking in a livable
environment for living creatures is a very challenging task.
Recently, a significant step toward this direction was reported:  a
microwave unidirectional cloak for a polarized wave in air
successfully hid a free-standing object along a single direction
with almost ideal performance, whose size was about ten wavelengths
large\supercite{landy_directional}. However, making a large-scale
living creature invisible to human eyes has not yet been possible.

In fact, all these difficulties in implementation stem from the need
of bending light around the hidden object preserving its phase.
Indeed, the rays that go around the hidden object have longer
physical paths. In order to preserve the phase, they need to travel
much faster than other rays going straightly in the external
environment, which subsequently leads to superluminal phase
propagation and extreme material
parameters\supercite{pendry,schurig,baile_cloak_review}. The
requirement of phase preservation is necessary at microwave
frequencies because the phase of microwaves can be easily detected
with an antenna. However, the premise of preserving the phase of
light in the natural light optical cloaking device lacks rationale.
Indeed, natural light is essentially randomly polarized and
incoherent and its phase is not well defined. Living creatures
cannot sense the phase of light and most of them, like humans, are
largely insensitive to polarization. Therefore, abandoning the
requirement of phase preservation for natural light cloaking opens
the door to hide large-scale living creatures.

Here we demonstrate in experiment that by abandoning the phase
preservation requirement it is possible to create invisibility
cloaking for natural light in multiple observation angles (see
supplementary videos documenting real life cloaking performance).
Such a cloak will act as a cloaking device operational on the ray
optics approximation. It will disregard the fine effects of
interference seen in wave optics but will offer good performance for
hiding macroscopic objects much larger than the wavelength of light.
Recent theoretical development in cloaking from non-Euclidean
transformation\supercite{leonhardt_broad,leonhardt_nonsuperluminal}
has provided an alternative way of abolishing phase preservation by
incorporating anisotropic materials. Our experimental demonstration
with only isotropic materials in incoherent light can serve as the
first simplified test of phase non-preservation cloak. Compared with
the previous design of unidirectional
cloak\supercite{xisheng,landy_directional}, our method can be easily
extended into multiple directions for arbitrary polarization at
broad range of optical frequencies, and therefore, can significantly
simplify the construction of a cloak in many real applications where
only a certain number of detectors or observers are involved. Using
the widely available optical glass we constructed polarization
insensitive cloaks hiding a fish in the fish tank and a cat in the
environment of human habitat.

\noindent\textbf{Results}

We start with analysis on how to simplify a perfect cloaking device
designed by the transformation optics approach proposed by Pendry
\emph{et al.}\supercite{pendry}. For comparison, we first analyze the
case when phase preservation is maintained, and later we will
abolish the phase preservation requirement. Consider a perfect
square cloaking device (Fig. 1a) with a square ``hole" at the center
opened from coordinate transformation. In the regions marked in
green in Fig. 1a, the wave needs to propagate with infinite phase
velocity. (See more details in Supplementary Information.) Here the
vertical coordinates (dotted red lines) represent wavefronts for the
rays going around the cloaked area. In this special case of square
cloak, we can make wavefronts perpendicular to rays everywhere in
the cloaking device even if the phase of outgoing waves is
preserved. As a result the cloak can be constructed without the need
for metamaterials, using only isotropic and nonsingular medium (Fig.
1b). The requirement for infinite phase velocity is therefore
removed. Because of its four-fold rotational symmetry, this
simplified cloak can work for four different incident directions
with arbitrary polarization.

The above example demonstrates the possibility to simplify a perfect
cloaking device using isotropic materials. However, because the
phase preservation is still maintained, superluminal phase
propagation is still required in the design unless the ambient
medium possesses a refractive index higher than unit (such as
water). Therefore, firstly, to achieve a broadband cloaking device
in the air environment, we have to abolish the phase preservation
requirement. Secondly, extending the cloaking performance into more
directions (\emph{e.g.} six directions) using isotropic materials is
almost impossible if phase preservation is required. In what follows
we will explore a more complex hexagonal cloak to demonstrate
cloaking without phase preservation in ambient water and air.

Fig. 1c shows a perfect, phase preserving omnidirectional hexagonal
cloak with extreme and anisotropic material parameters designed with
the transformation optics approach, which can be constructed using
metamaterials only. In the simplified cloak, when abolishing the
phase preservation requirement, we still expect the incident rays to
return to their original trajectories. Fig. 1d shows that the phase
rearrangement causes wavefront dislocations in the simplified cloak.
However, as will be demonstrated below, distortions caused by them
can be disregarded  in experimental observations.

Using the simplified six-fold cloaking design, we first construct a
cloak hiding a fish in the aquatic environment using the widely
available optical glass. Fig. 2a shows the experimental setup where
a cloak is immersed in a fish tank filled with water as a living
environment for a goldfish. The cloak is constructed with six pieces
of glass with $n=1.78$ placed in a hexagonal hollow transparent
container with negligible thickness. A camera placed in front of the
tank records the dynamic scene from the front observation angle.
Figs. 2b-e show the dynamic process of the goldfish swimming from
inside the cloak to the external aquatic environment. When swimming
inside the cloak, the goldfish becomes invisible and does not block
the scene of green plants behind the cloak. The edges of the
transparent container are still visible because of some glue residue
left at the edges of this homemade container. This cloak works for
six different incident directions.

Now we proceed to extend the simplified cloak from the aquatic
environment to a terrestrial one with air as ambient medium. Here we
could construct a four-directional cloak by using design in Fig. 1b.
However, for simplicity of demonstration, we construct a
unidirectional cloak instead that can hide terrestrial creatures
from forward and backward observation angles. We use a cat as the
terrestrial creature in the experiment. The cloak is constructed
with a few pieces of the same glass material with $n=1.78$, as
indicated in dark blue in Fig. 3a. To demonstrate decisively the
capability of this cloaking strategy in hiding creatures especially
in a dynamic background in natural light illumination, we use an
office projector equipped with an incandescent bulb to project a
dynamic field scenery through the cloak. The light emitted from the
incandescent bulb has very similar characteristics to natural light
in terms of random polarization, incoherence, and continuous visible
spectrum. A screen behind the cloak is used to display the projected
image. Video recording illustrated the cloak performance can be
found in Supplementary Information.

Figs. 3b-d show pictures on the screen captured by a
digital camera behind the screen. Fig. 3b is the picture when there
is just the cloak in front of the background scenery where a yellow
butterfly is flitting around flowers. The cloak casts some moderate
edge shadows due to the fact that the light coming from the
projector is divergent rather than being ideally parallel. In Fig.
3c, a living cat is stepping into the cloak. One can see clearly
that the head and forelegs of the cat inside the cloak become
invisible. In Fig. 3d, the main body of the cat has settled inside
the cloak and becomes invisible, while the head and forelegs of the
cat outside the cloak are still visible and block the white flower
in the middle. Particularly, at this moment, the butterfly is
flitting quickly from the upper left to the lower right behind the
cloak, but still can be seen clearly on the screen through the body
of the cat.

\noindent\textbf{Discussion}

Although the demonstrated cloaking solution is only effective for
several observation directions, our work has successfully made a
step toward practical application of invisibility cloaking in hiding
large-scale creatures in plain sight. By reconfiguring the prisms
the cloak operators could make them disappear from the sight along
any given direction which lands to important security, entertainment,
and surveillance applications.

\newpage

\noindent\textbf{Acknowledgements}
\\This work was sponsored by the
National Natural Science Foundation of China under Grants No.
61275183 and No. 60990322, the Foundation for the Author of National
Excellent Doctoral Dissertation of China under Grant No. 200950, the
Fundamental Research Funds for the Central Universities under Grant
No. 2011QNA5020, the Chinese Scholarship Council Foundation under
Grant No. 2011833070, the Program for New Century Excellent Talents
(NCET-12-0489) in University, Nanyang Technological
University(M4080806, M4081153) and the Singapore Ministry of
Education under Grant No. MOE2011-T3-1-005.

\noindent\textbf{Author Contributions}
\\H.C. conceived the original idea. B. Zhang provided the explanation on incoherence. H.C., H.W., B. Zheng, and B. Zhang designed the cloaks.  H.C., B. Zhang, and X.Z designed and supervised the experiments. B. Zheng and L.S. carried out experiments. B. Zhang, H.C., and N.Z originated the manuscript and supplementary materials and interpreted the results. All authors joined discussion and reviewed the manuscript.

\noindent\textbf{Competing Financial Interests statement}
\\The authors declare no competing financial interests.

\noindent \textbf{Links to the supplementary videos:}\\
https://www.dropbox.com/s/6kslerc5rpyek38/ChenS1.mov\\
https://www.dropbox.com/s/4tw4jg1hr9mh0zk/ChenS2.mov

\newpage

\begin{figure}
\begin{centering}
\includegraphics[width=1\columnwidth,draft=false]{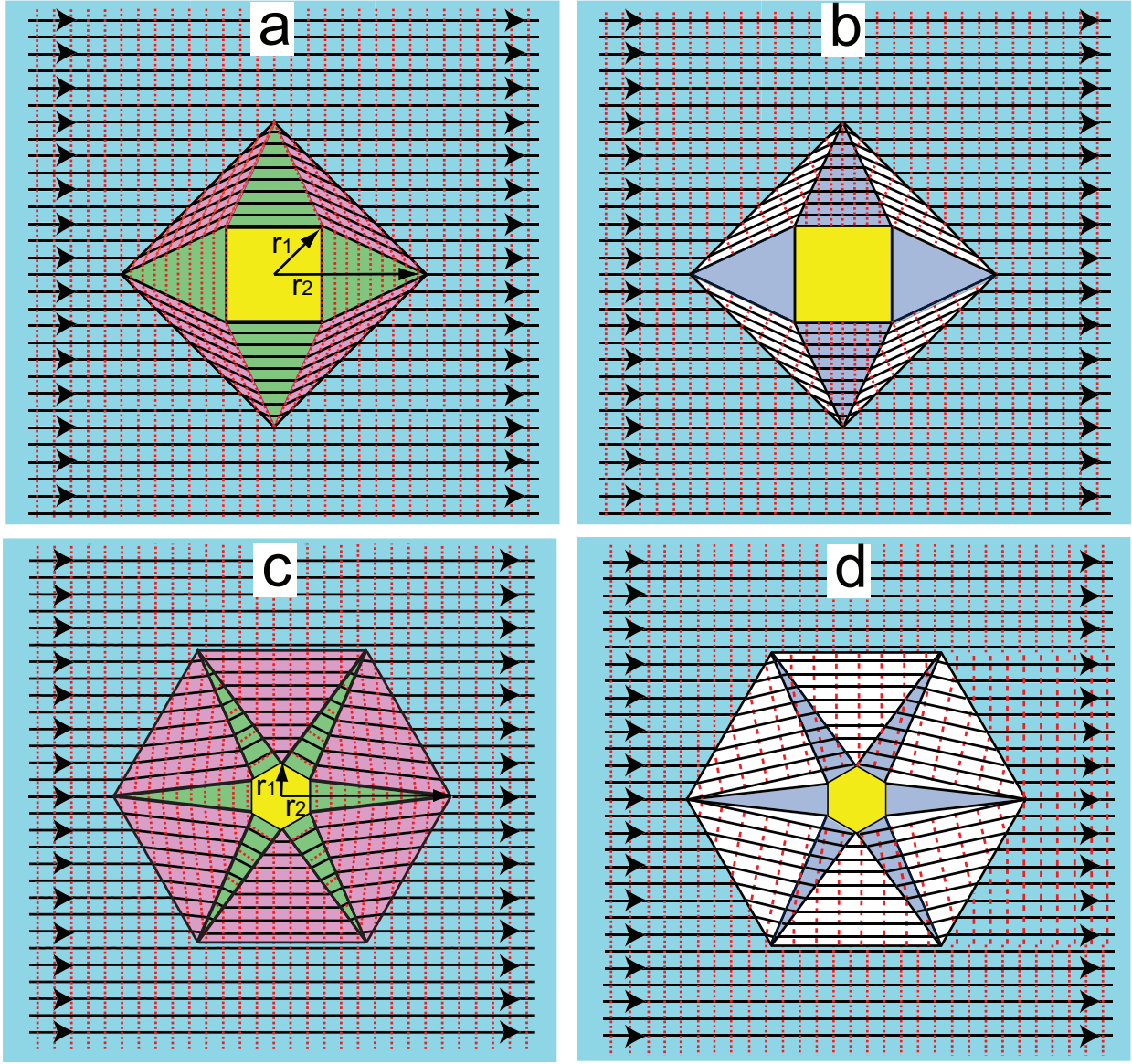}% Here is how to import EPS art
\caption{\label{fig:figure1}  \textbf{Principle of natural light
cloak}. Horizontal rays are incident from left to right. Red dotted
vertical lines represent wavefronts when illumination is coherent.
The central hidden region is marked with yellow. \textbf{a}, A
perfect square cloak designed by applying a coordinate
transformation to open a square ``hole" at the center. The wave has
to propagate with infinite phase velocity in the singular
anisotropic region (marked in green). \textbf{b}, A four-directional
square cloak with incident rays propagating horizontally. Wavefronts
are perpendicular to rays. \textbf{c}, A perfect hexagonal cloak
designed from coordinate transformation. \textbf{d}, A
six-directional hexagonal cloak with incident rays propagating
horizontally. Wavefronts are perpendicular to rays. In \textbf{a}
and \textbf{c}, the cloaks require extreme and anisotropic material
parameters. In \textbf{b} and \textbf{d}, the cloaks can be greatly
simplified with isotropic materials while the invisibility
performance is maintained in multiple directions.}
\end{centering}
\end{figure}

\begin{figure}
\includegraphics[width=1\columnwidth,draft=false]{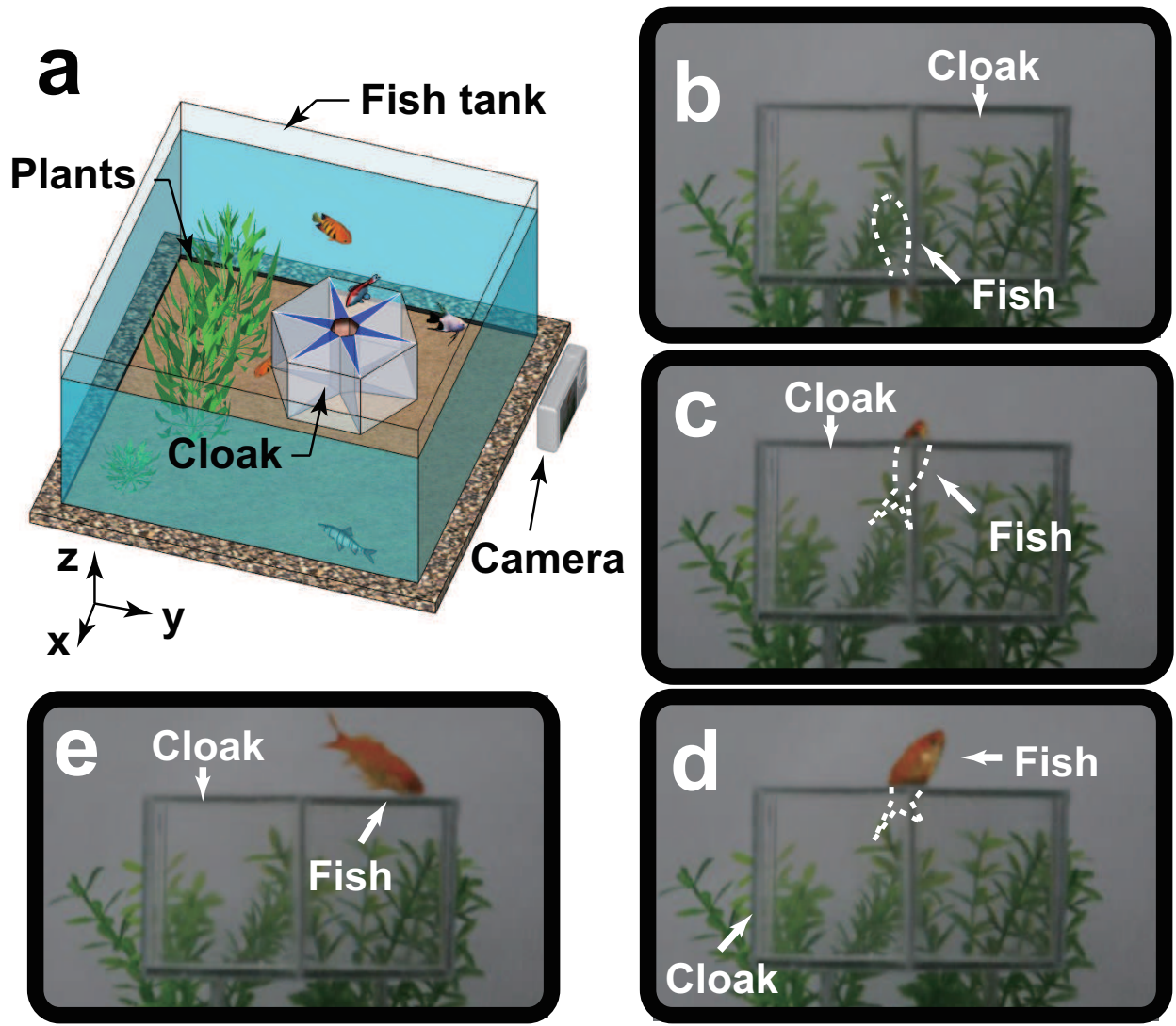}% Here is how to import EPS art
\caption{\label{fig:figure2} \textbf{Experimental observation of
fish in aquatic ray cloak}. \textbf{a}, Experimental setup in a fish tank.
The cloak is constructed with six pieces of glass with $n=1.78$
(indicated in dark blue) enclosed in a hollow hexagonal container.
\textbf{b-e}, Dynamic monitoring of a fish swimming through the aquatic ray
cloak. The outline of the invisible fish body is indicated by dotted
lines. \textbf{b}, The main fish body inside the cloak is invisible but only
the tail outside of the cloak is visible. \textbf{c}, Only the fish head
outside of the cloak is visible. \textbf{d}, The main body of the fish comes
out of the cloak and thus becomes visible. \textbf{e}, The whole fish has
come out from the cloak.}
\end{figure}

\begin{figure}
\includegraphics[width=1\columnwidth,draft=false]{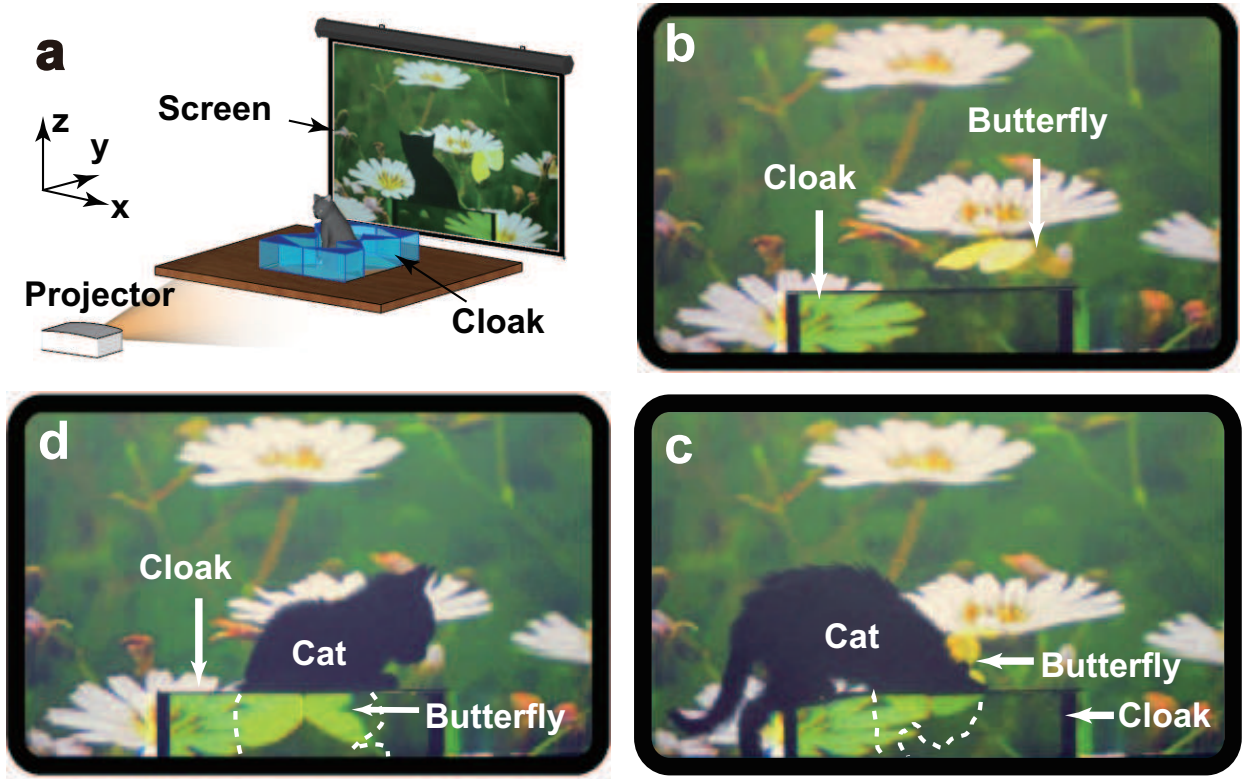}% Here is how to import EPS art
\caption{\label{fig:figure3}  \textbf{Experimental observation of
cat in terrestrial ray cloak}. \textbf{a}, Experimental setup to
test the cloaking performance.  The cloak in dark blue is
constructed with glass ($n=1.78$) and has dimensions of 0.3 m long
(along the \emph{y} direction), 0.26 m wide (along the \emph{x}
direction), and 0.07 m high (along the \emph{z} direction). An
office projector projects a movie through the cloak onto the screen
behind the cloak. A camera (not shown here) placed behind the screen
records the movie on the screen. A live cat is sitting inside the
cloak. \textbf{b-d},The images displayed on the screen when
(\textbf{b}) only the cloak is present, (\textbf{c}) a live cat is
stepping into the cloak, and (\textbf{d}) the cat's main body has
settled inside the cloak, respectively. The outline of the invisible
body of the cat is indicated by dotted lines. During the whole
process, the butterfly in the background scenery is flitting about.
}
\end{figure}

\end{document}